\documentclass[a4paper,fleqn]{cas-dc}

\usepackage[authoryear,longnamesfirst]{natbib}
\usepackage{bm}
\usepackage{amsmath}
\usepackage{subfigure}
\usepackage{amssymb}
\usepackage{graphicx}
\usepackage{textcomp}
\usepackage{epsfig}
\usepackage{psfrag}
\usepackage{subfigure}
\usepackage{epstopdf}
\def\tsc#1{\csdef{#1}{\textsc{\lowercase{#1}}\xspace}}
\tsc{WGM}
\tsc{QE}

\begin{document}
\let\WriteBookmarks\relax
\def\floatpagepagefraction{1}
\def\textpagefraction{.001}

\shorttitle{Simultaneously exploring multi-scale and asymmetric EEG features for emotion recognition}

\shortauthors{Wu et al.}

\title [mode = title]{Simultaneously exploring multi-scale and asymmetric EEG features for emotion recognition}



%

\author[1]{Yihan Wu}

\author[1]{Min Xia}

\author[1]{Li Nie}

\author[1,3]{Yangsong Zhang}
\cormark[1]
\ead{zhangysacademy@gmail.com}

\author[2]{Andong Fan}
\cormark[1]
\ead{fad@cdut.edu.cn}

\affiliation[1]{organization={School of Computer Science and Technology, Laboratory for Brain Science and Medical
Artificial Intelligence, Southwest University of Science and Technology},
            city={Mianyang},
            postcode={621010},
            country={China}}

\affiliation[2]{organization={College of Mathematics and Physics, Chengdu University of Technology},
        	city={Chengdu},
        	postcode={610059},
        	country={China}}

 \affiliation[3]{organization={MOE Key Lab for Neuroinformation, University of Electronic Science and Technology of China},
        	city={Chengdu},
        	country={China}}

\cortext[cor1]{Corresponding author}

\begin{abstract}
In recent years, emotion recognition based on electroencephalography (EEG) has received growing interests in the brain-computer interaction (BCI) field. The neuroscience researches indicate that the left and right brain hemispheres demonstrate activity differences under different emotional activities, which could be an important principle for designing deep learning (DL) model for emotion recognition. Besides, owing to the nonstationarity of EEG signals, using convolution kernels of a single size may not sufficiently extract the abundant features for EEG classification tasks. Based on these two angles, we proposed a model termed  Multi-Scales Bi-hemispheric Asymmetric Model (MSBAM) based on convolutional neural network (CNN) structure. Evaluated on the public DEAP and DREAMER datasets, MSBAM achieved over 99\% accuracy for the two-class classification of low-level and high-level states in each of four emotional dimensions, i.e., arousal, valence, dominance and liking, respectively. This study further demonstrated the promising potential to design the DL model from the multi-scale characteristics of the EEG data and the neural mechanisms of the emotion cognition.
\end{abstract}


\begin{highlights}
\item Simultaneously using multi-scale and bi-hemispheric asymmetric features is beneficial to recognize emotional states.
\item The proposed method MSBAM yield better performance than the compared baseline methods on DEAP dataset and DREAMER dataset.
\item MSBAM can also achieve satisfactory performance on four-class classification task.
\end{highlights}


\maketitle

\section{Introduction}\label{sec:introduction}
Emotion is a kind of physiological and psychological phenomenon playing a significant role in daily life\cite{Dolan1191}\cite{2003What}. In recent years, emotion recognition based on machine learning and deep learning(DL) methods has attracted growing attention of the research community~\cite{torres2020eeg}. Emotions can be detected from different modalities, such as facial expression~\cite{Zhaosr2021}, speech~\cite{ELAYADI2011}, and physiological data~\cite{Lipy2019}, etc. Among these, the physiological signals are hard to fake and show a decided advantage in measurement of spontaneous mental activity under different emotional states.

Human physiological signals can be measured by different imaging modalities, such as functional magnetic resonance	imaging (fMRI), magnetoencephalography (MEG), functional near-infrared spectroscopy (fNIRS), and electroencephalography (EEG), et al. Benefiting from properties such as high temporal resolution, non-invasiveness, low cost, and the portability of devices, EEG has been widely employed in the field of emotion recognition~\cite{liwei2021}.

In the field of emotion recognition, research objects can be defined in many ways. From the emotional modeling point of view, common theories include discrete emotional model and emotional circumplex model, which correspond to different evaluation criteria. For instance, Berke K{\i}l{\i}{\c{c}} and Serap Ayd{\i}n proposed a method based on support vector machine (SVM) and graph theoretical network measures to classify five pairs of discrete emotions. Their method achieved the best accuracy of 80.65\%~\cite{kilicc2022classification}. In addition, some researchers tried to explore other related indicators using the emotional EEG. For example, Serap Ayd{\i}n proposed an emotional complexity marker, i.e. phase space trajectory matrix. They employed this marker to classify the gender of subjects and distinguish nine emotional states from baseline state~\cite{8933102}. In this study, we mainly focus on discussing the classification between low-level and high-level of the evaluative dimensions in the emotional circumplex model.

To differentiate emotional states, the conventional algorithms usually include a feature extractor and a classifier. Various hand-crafted features have been employed to extract the differences between different emotional states. For instance, Wen et al. utilized Pearson correlation coefficient (PCC) to estimate the correlation between all channel pairs. They extracted the PCC feature and convolutional neural networks (CNN) features parallelly to classify the emotional states. Their method achieved  average accuracies of 77.98\% and 72.98\% in valance and arousal of the DEAP dataset, respectively~\cite{2017A}. Zheng et al. extracted five hand-crafted features, i.e. power spectral density (PSD), differential entropy (DE), differential asymmetry, rational asymmetry, asymmetry and differential causality features, to recognize the emotion using SVM and graph regularized extreme learning machine (GELM) classifier, respectively. The DE features and GELM classifier obtained 69.67\% accuracy for four classification task on valance-arousal space~\cite{2017Identifying}. Moon et al. adopted the PCC, phase-locking value~(PLV) and transfer entropy (TE) to calculate the hand-crafted features, and employed CNN to extract advanced features to classify the emotional states. Their method achieved 87.75\% accuracy in valance dimension with the data length of 3 seconds~\cite{MOON202096}.

With the rapid development of deep learning, increasing researchers are pursuing end-to-end solutions to replace the conventional classification methods based on handcraft features. Yang et al. proposed a parallel convolutional recurrent neural network model by combining CNN and Long-Short Term Memories neural network(LSTM). The CNN is adopted to extract the inter-channel correlation, and the LSTM is designed to mine the temporal contextual correlation. Their method  achieved 90.80\% and 91.03\% performance on valance and arousal of DEAP with data length of 1 second, respectively~\cite{2018Emotion}. Ma et al. proposed a method that applied residual structure to LSTM. Multimodal signal were input into two residual LSTM networks that are shared part weight to extract the emotion related high-level features. This method obtained 92.30\% and 92.87\% accuracies on the valance and arousal dimensions of DEAP dataset with data length of 1 second, respectively~\cite{ma2019emotion}.  Yin et al. proposed a method by fusing graph convolutional neural networks (GCNN) and LSTM. Several GCNNs were employed to extract features of graph domain. LSTM are used to extract the temporal features and detect their changes. This method obtained 90.45\% and 90.60\% accuracy on DEAP with data length of 6 seconds, respectively~\cite{YIN2021106954}.

For the past few years, increasing studies have started to design the DL models by considering the physiological mechanisms of emotions. Bi-hemispheric discrepancy under different emotion states is a vital neural mechanism, which has been used in several studies with DL models for emotion recognition~\cite{2018A,2021Differences}. For instance, Li et al. proposed a method using LSTM to capture the high-level discrepancy features from bi-hemispheric hand-crafted features. They input those features into a domain adaption model to classify the emotional states. This model yielded 92.38\% and 84.14\% accuracy on the subject dependent and independent task on the SEED dataset, respectively~\cite{2018A}. In another study, Li et al. proposed a model, termed R2G-STNN,  to extract regional features according to the spatial region concerning physiological function, and fused the regional and global features. This method achieved 93.38\% and 84.16\% accuracy on the SEED dataset~\cite{2019From}.

EEG signals are nonstationary, using convolution kernels of a single size may not sufficiently extract the abundant features for EEG classification tasks. Previous studies have demonstrate that using various convolution kernels of different sizes could learn multi-scale EEG features that are beneficial for different EEG classification tasks~\cite{Ko2021}. For example, Li et al. proposed a multi-scale fusion CNN model based on attention mechanism for motor imagery classification~\cite{lidl2020}. The experimental results indicated that the model achieved a better performance compared with the baseline methods. The multi-scale CNN has also been introduced in emotion recognition. Phan et al. proposed a 2D CNN model with convolution kernels of different sizes for arousal and valence binary classification. They used kernel size of $5\times5$ and $7\times7$ to extract spatial features to describe the short-range and long-range relations between EEG channels. This method achieved 98.27\% and 98.36\% accuracies on the valance and arousal dimensions of DEAP dataset~\cite{phan2021eeg}.

However, the physiological mechanisms and multi-scale features of EEG have not been simultaneously considered in a DL model for emotion recognition. Based on this consideration, we proposed a DL model termed Multi-Scales Bi-hemispheric Asymmetric Model (MSBAM) to explore the multi-scale asymmetric information. MSBAM is composed of the parallel spatial domain feature extractor and temporal domain feature extractor, followed by a fully connected layer classifier. We conducted extensive experiments on the public DEAP dataset and DREAMER dataset to evaluate the performance of MSBAM and the compared baseline methods. The results indicate that our MSBAM achieve better performance than the baseline methods.

The remainder of this paper is organized as follows. Section 2 introduces materials and methods. Section 3 describes the settings and results of extensive experiments, comparisons between MSBAM and the baseline methods are also be provided in this section. Section 4 and 5 present the discussions and conclusion.

\section{Materials and Methods}
\subsection{Datasets}
The public DEAP and DREAMER dataset were adopted to validate the performance of our MSBAM, which are widely used for emotion recognition researches.

The DEAP dataset is a multimodal dataset presented by Koelstra et al.~\cite{Koelstra2012DEAP}. Thirty-two healthy subjects participated in the experiment. They were watching 40 one-minute pieces of music videos. The EEG and peripheral physiological signals were recorded when they were watching videos. Forty electrodes (32 for EEG and 8 for peripheral physiological signals) were used for the EEG recording. Participants were asked to rate a score for each video from 1 to 9 to evaluate the levels of four emotional dimensions, i.e., arousal, valence, dominance and liking, respectively. Each trial contains 3 s baseline data and 60 s task data. For the offline analysis, the EEG signals were first downsampled to 128 Hz from 512 Hz, and re-referenced to the common average reference. Electrooculogram (EOG) artifacts were removed. Then, the EEG data was filtered by a bandpass filter with 4-45 Hz. The details can be found in the reference~\cite{Koelstra2012DEAP}.

The DREAMER dataset was presented by Katsigiannis et al.~\cite{7887697}. Twenty-three subjects (9 females and 14 males) were recruited in the experiments. The dataset was collected with 14 electrodes at a sampling rate of 128 Hz during the subjects watching 18 emotional film clips. The electrode placement was according to the international 10-20 system. The duration of the film clips are between 65 and 393 seconds. Before watching each emotional film clip, subjects are asked to watch a neutral film clip that is regarded as having no valance to record the baseline signals. After the EEG experiments, the subjects were asked to rate a score for each clip from 1 to 5 for arousal, valence and dominance by self-assessment, respectively.

\subsection{The classification tasks for emotional states }
In this work, MSBAM was designed to distinguish between low-level and high-level emotional states according to the circumplex model with the EEG data. The classification tasks were conducted independently in each of the four emotion dimensions, i.e., arousal, valence, dominance and liking. To generate the two-class data in each emotion dimension, a threshold of self-rating scores was chosen to label each trial with a low-level or high-level emotional state. For example, in the valance dimension, supposing the threshold was 5, the trials with self-rating score higher than 5 were labeled as high valance, other trials were labeled as low valance. The values of the threshold for the DEAP and DREAMER datasets were different according to previous studies. Specifically, for the four dimensions of the DEAP dataset, the threshold was set as 5 as in the previous studies~\cite{liu2016emotion, 9321553}. Similarly, for the DREAMER dataset, the threshold was set as 3 because the score range is from 1 to 5~\cite{9321553,cui2020eeg}.

\subsection{Data processing}

\begin{figure}[ht]
	\centering
	\includegraphics[scale=0.25]{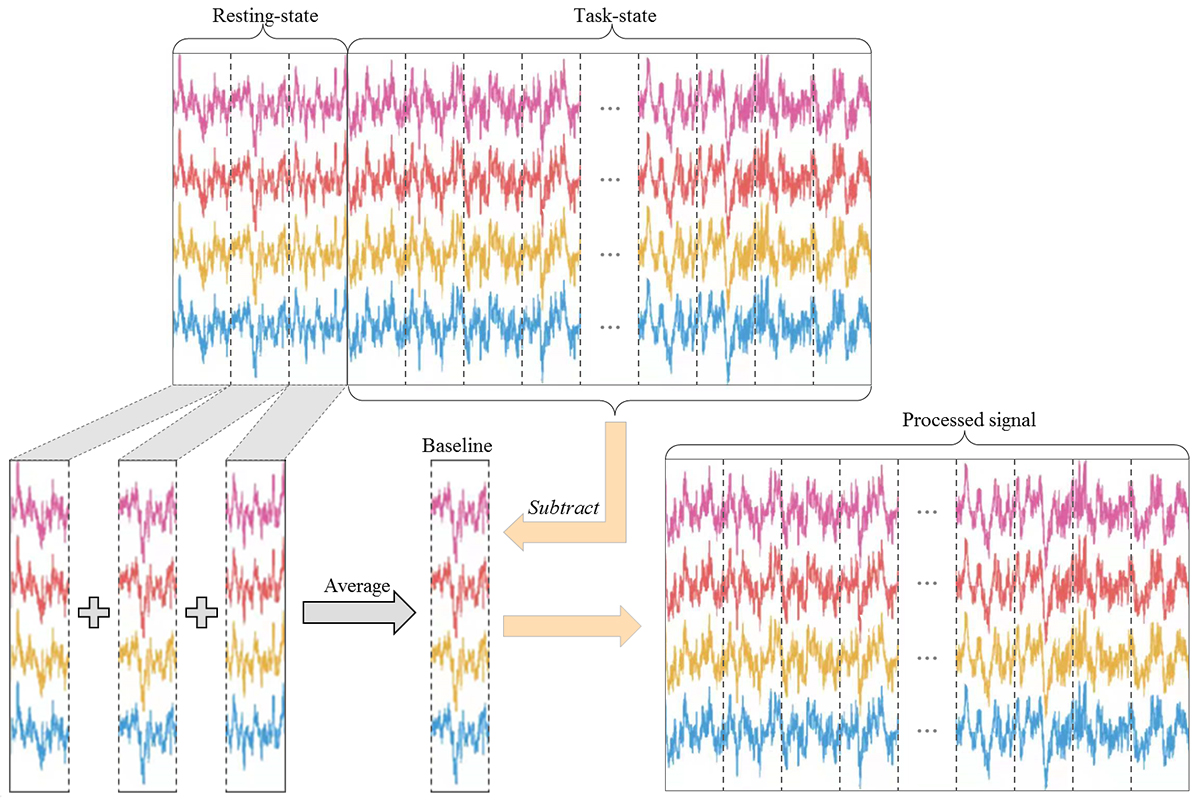}
	\caption{The diagram of baseline correction. The data of the first 3 s in each trial of the DEAP dataset is regarded as resting-state data, and the baseline data of last 3 s in each trial of the DREAMER dataset . All the task-state data should subtract the average signal of resting-state second by second.}
	\label{Figure1}
\end{figure}

In order to reduce the noise and improve the stability of the EEG data, baseline correction and z-score normalization were implemented on the two datasets, which were commonly pre-processing procedures for EEG~\cite{2021Differences, cui2020eeg}. Previous studies demonstrated that these operations can improve emotion recognition accuracies by reducing the interference of basic emotional state before the task period~\cite{2018Emotion}. The diagram of baseline correction is illustrated in the Fig.~\ref{Figure1}.

Take for instance the DEAP dataset. First, the 3-s baseline data in each trial were divided into three segments of 1 s length without overlap. These segments were further averaged to obtain the baseline signal of 1 s length, called resting-state segment. Second, the 60 s task data in each trial were divided into 60 segments of 1 s length without overlap, termed task-state segments. Third, each task-state segment subtracted the resting-state segment.  After these steps, we concatenated all the processed task-state segments to construct the new task-state data of 60 s. At the last, z-score normalization was utilized on each channel of the task-state data. For the DREAMER dataset, we selected the baseline data of the last 3 s in each trial to calculate the resting-state segment for each trial.

After the pre-processing procedure, the new task-state data will be segmented with \emph{Wnd} seconds window without overlap. Finally, we obtained samples $D\in\mathbb{R}^{C \times T}$, where $C$ denotes the number of channels and $T = Fs * Wnd$ represents the number of sample points. $Fs$ denotes the sampling rate of 128 Hz. For the parameter $Wnd$, we set it to be 1 in the following experiments as that in the previous studies~\cite{2021Differences, ma2019emotion, 9321553}.

Overall, we have obtained 32(subjects) $\times$ 40(trials) $\times$ 60(segments per trial) =76800 samples in total for DEAP dataset. Similarly, 23(subjects) $\times$ 3728(segments per subject) = 85744 samples are obtained for DREAMER dataset.

\subsection{3D representation of the EEG data}

\begin{figure}[!htb]
	\centering
	\includegraphics[scale=0.24]{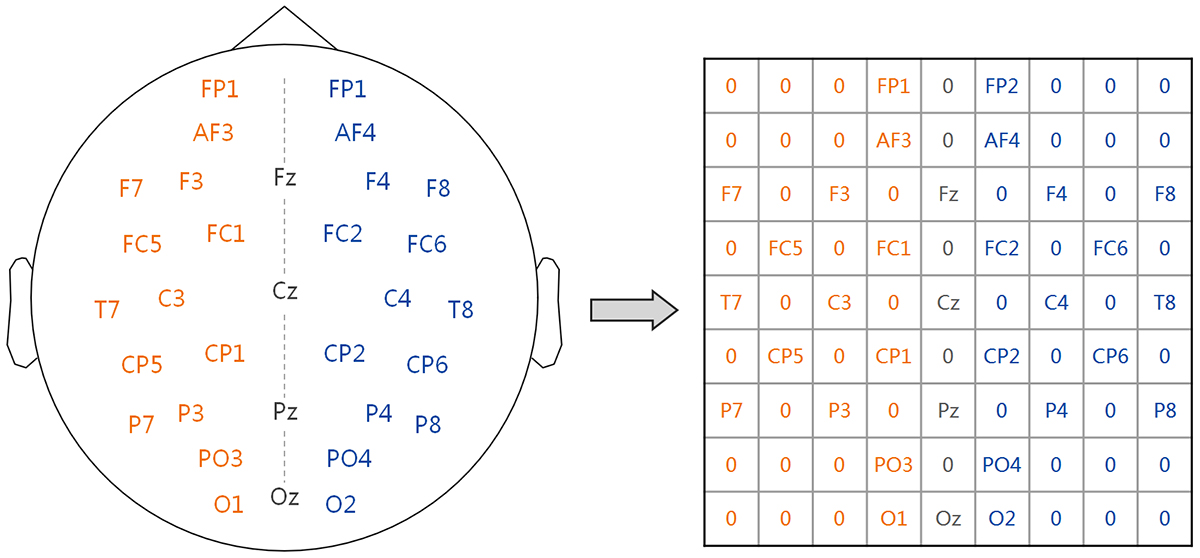}
	\caption{A schematic diagram of a 2D representation of EEG channel locations with a 9 $\times$ 9 matrix at each sample point. }
	\label{Figure2}
\end{figure}

Traditionally, EEG data is represented as 2D matrix with the shape of channels $\times$ sample points for most algorithms. Under the 2D representation, the data from all used channels at a sampling time point are arranged into a column vector, the topological information among different channels would be lost. In recent years, the 3D representation has been introduced for EEG data~\cite{zhao2019multi}. In this way, the data from all used channels at one sample point are arranged into a 2D matrix according to distribution of the international 10-20 system, which can retain the spatial information among channels to some extent. We denote this operation as spatial transformation. The schematic diagram is shown in Fig.~\ref {Figure2}. We implemented the spatial transformation at all sample points to obtain the 3D representation of the EEG data. This 3D representation can preserve both temporal information and spatial information of EEG data, which has been adopted for various EEG classification tasks~\cite{cui2020eeg,zhang2021atten, zhao2019multi}.

As the input of our model, each sample $D\in\mathbb{R}^{C \times T}$ is transformed to a 3D spatial-temporal matrix $D\in\mathbb{R}^{H \times W \times T}$. In current study, $C=$ 32 and 14 for DEAP and DREAMER respectively. $T=128$, and $H$ and $W$ were set to 9. Then, each sample was in a 3D shape of size 9 $\times$ 9 $\times$ 128.

\subsection{The construction of proposed MSBAM}
\begin{figure}[!htb]
	\centering
	\includegraphics[scale=0.25]{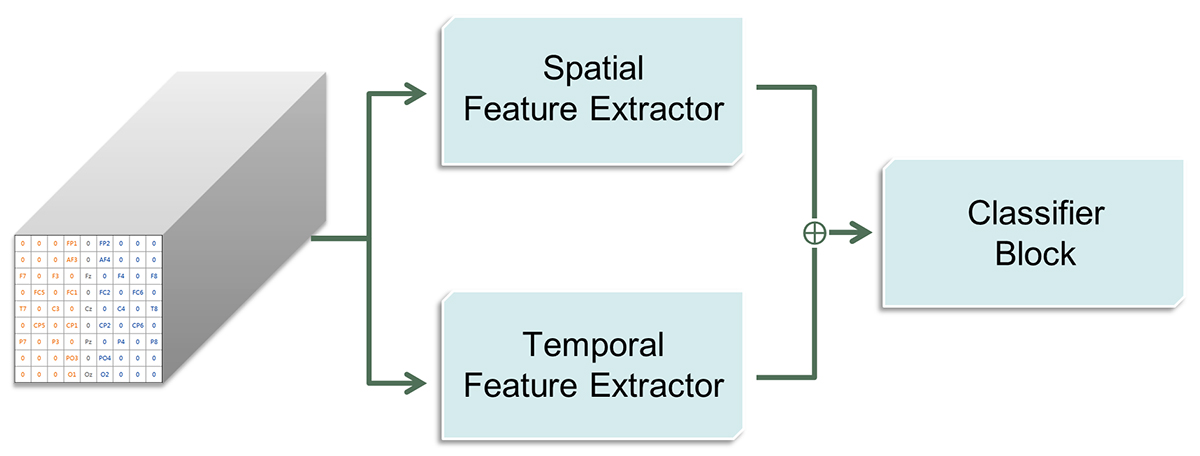}
	\caption{The structure of MSBAM.}
	\label{Structure}
\end{figure}

Previous studies indicate the significance of physiological mechanisms and multi-scale features of EEG for emotion recognition~\cite{2018A,phan2021eeg}. However, the correlations and differences of asymmetric features in various time scales have not been considered in a DL model for emotion recognition. Based on this consideration, our MSBAM contains three parts, i.e., a spatial feature extractor block, a temporal feature extractor block, and a feature classification block. The structure and details of MSBAM are shown in Figs.~\ref{Structure},~\ref{Spatial} and~\ref{MSBAM}, respectively.

\subsubsection{Spatial feature extractor block}
\begin{figure*}[!htb]
	\centering
	\includegraphics[scale=0.45]{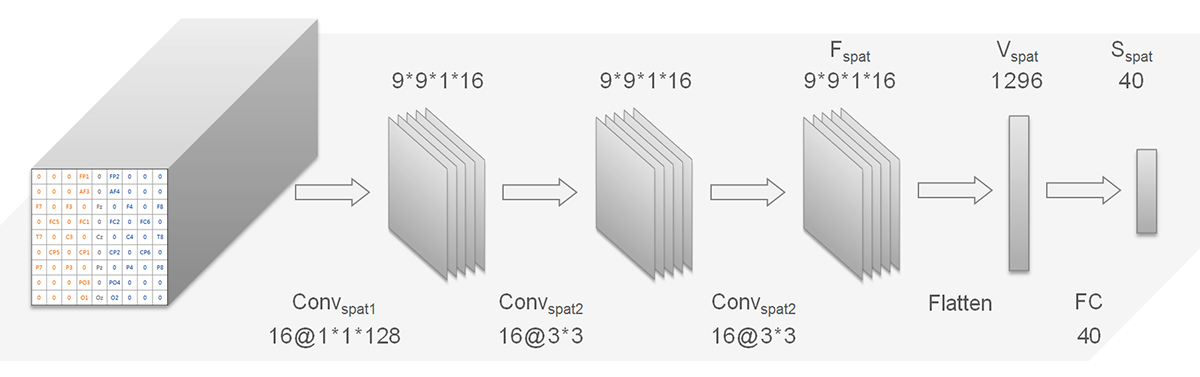}
	\caption{A schematic diagram of overall structure of the spatial feature extractor block. }
	\label{Spatial}
\end{figure*}

For the spatial feature extractor block, as shown in Fig.~\ref {Spatial}, it contains three convolution(Conv) operations and one fully connected layer (FC). The first Conv operation, denoted as $Conv_{spat1}$, contains a convolution ($conv_1$) with 16 kernels of size 1 $\times$ 1 $\times$ 128, and a Scaled Exponential Linear Units (SELU) activation function, which is expressed as:

\begin{equation}
	{Conv_{spat1}(\cdot) = \sigma(conv_1(\cdot))}
\end{equation}
where $ \sigma $ denotes the SELU function. This procedure is able to compress the data from 3 D into 2 D, and all values in the time dimension are transformed to single feature value. The remaining two Conv operations are similar to the $Conv_{spat1}$ except for the convolution with 32 kernels of size 3 $\times$ 3 and padding size of (1, 1). Through these Conv operations successively, a feature $ F_{spat}  \in \mathbb{R}^{9 \times 9 \times 1 \times 16}$ is obtained to represent the spatial feature of the EEG data.

Then the feature $F_{spat}$ is flattened to a vector and input into a FC layer followed by a batch normalization layer and softmax activation to obtain the normalized feature \emph{$S_{spat}$}, which could be described as:

\begin{eqnarray}
	&V_{spat} &= Flatten(F_{spat})\\
	&V_{fc} &= BN( W \cdot V_{spat} + b)\label{linear_softmax_1a}\\
	&&= [V_1, V_2, ..., V_{40}]  \in \mathbb{R}^{40} \label{linear_softmax_1b}\\
	&\bar{V}_i &= \frac{exp(V_i)}{\sum_{k = 1}^{40}exp(V_k)}, i = 1, 2,..., 40  \label{linear_softmax_2} \\
	&S_{spat} &= [\bar{V}_1, \bar{V}_2,..., \bar{V}_{40}]\in \mathbb{R}^{40} \label{linear_softmax_3}
\end{eqnarray}
where \emph{W} is the weight matrix, \emph{b} is the bias. The procedures in the formulas (\ref{linear_softmax_1a})-(\ref{linear_softmax_3}) are denoted as the $Feature Normali\-zation Layer (FNL)$, and will be utilized in the temporal feature extractor block.

\subsubsection{Temporal feature extractor block}
\begin{figure*}[!htb]
	\centering
	\subfigure[]{\includegraphics[scale=0.25]{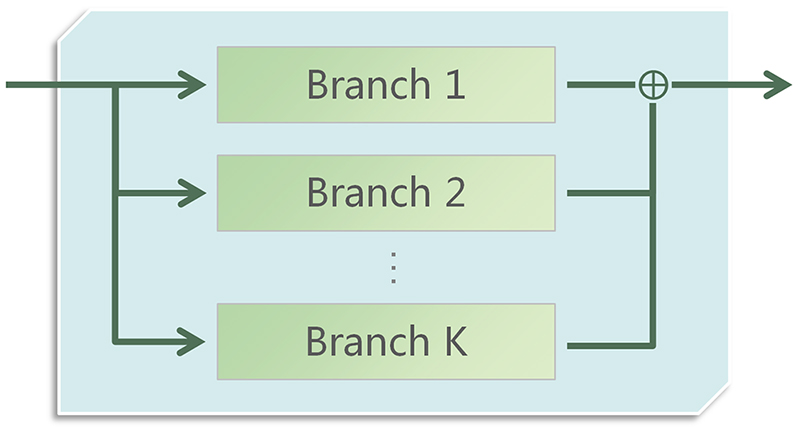}
		\label{Temporal}}
	\vfil
	\subfigure[]{\includegraphics[scale=0.33]{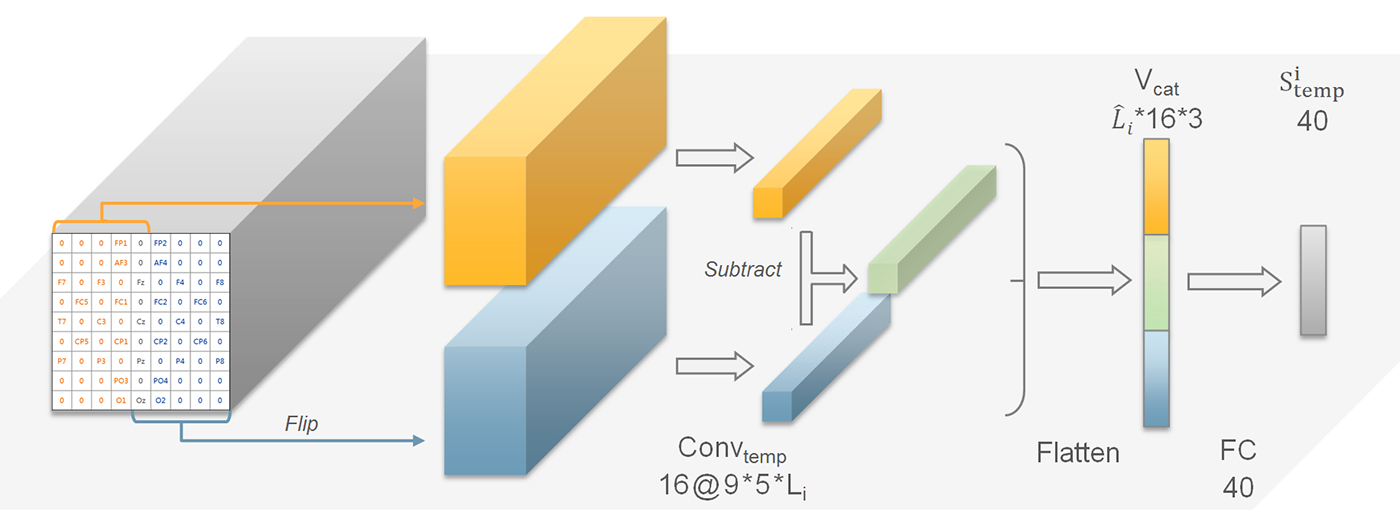}
		\label{Branch}}
	\caption{(a) shows overall structure of the temporal feature extractor block. $K$ denotes the number of branches, and K=2 is a proper configuration proved by our experiment. (b) shows the \emph{i}-th bi-hemispheric asymmetric feature extractor.}
	\label{MSBAM}
\end{figure*}

For the temporal feature extractor block, as is shown in Fig.~\ref {Temporal}, it contains $K$ branches, each of which contains a convolution operation named \emph{${Conv_{temp}^{k}}$}, a flatten-concatenation operation, and a FNL.
Multiple branches that using temporal convolution kernels of different lengths could learn multi-scale EEG features which is beneficial for emotion recognition tasks.

In each branch, the EEG data \emph{D} is split into two parts in spatial dimension. As shown in Fig.~\ref{Branch}, the first part, denoted as $D_l\in \mathbb{R}^{9 \times 5 \times 128}$, comes from the columns of [1, 2, 3, 4, 5] of the $9 \times 9$ matrix as shown in Fig.~\ref{Figure2}, which represents the data from left hemisphere. Similarly, the second part, denoted as $D_r \in \mathbb{R}^{9 \times 5 \times 128}$, comes from the columns of [5, 6, 7, 8, 9], which represents the data from right hemisphere. To remain the unified location of electrodes between $D_l$ and $D_r$, a horizontal flip is implemented on $D_r$, which rearranges the columns as [9, 8, 7, 6, 5]. After the process, $D_l$ and $D_r $ are first input into the same ${Conv_{temp}^{k}}$ block, which is illustrated in the following:
\begin{equation}
	{Conv_{temp}^{k}(\cdot) = \sigma(conv_k(\cdot))}
\end{equation}
where $ k $ denotes $k$-th branch, and $ \sigma $ is SELU activation function. The $conv_k$ conducts a 3D convolution with 16 kernels of size 9 $\times$ 5 $\times$ $L_i$ and the stride of size 9 $\times$ 5 $\times$ $S_i$. The \emph{$L_i$} is the time scale of the extractor, and the \emph{$S_i$} is equal to $L_i$//2.

%
After the ${Conv_{temp}}$ block, the \emph{$F_l$} and \emph{$F_r$} are obtained and utilized to calculate discrepant features, which are denoted as $F_a\in \mathbb{R}^{1 \times 1 \times \hat{L} \times 16}$. The calculation is described as follows:
\begin{eqnarray}    \label{F_a}
	F_a = F_l - F_r
\end{eqnarray}

Then, these three kinds of features, i.e., \emph{$F_l$}, \emph{$F_r$} and \emph{$F_a$}, are first flattened to vectors ($V_l\in \mathbb{R}^{\hat{L}\times 16}$, $V_r\in \mathbb{R}^{\hat{L} \times 16}$, and $V_a\in \mathbb{R}^{\hat{L} \times 16}$), and further combined into a vector $V_{cat}\in \mathbb{R}^{\hat{L} \times 16 \times 3}$. The feature $V_{cat}$ is input into FNL to achieve unified representation. We denoted the $i-th$ feature from $i-th$ branch as $S_{temp}^i\in \mathbb{R}^{40}$.

Finally, all the $K$ features from all the $K$ branches were concatenated as one vector $S_{temp}$:
\begin{equation}
	{S_{temp} = [S_{temp}^{1}\parallel S_{temp}^{2}\parallel ... \parallel S_{temp}^{K}]  \in \mathbb{R}^{40 \times K}} \label{cat_temp}
\end{equation}

After these procedures above, we obtain the asymmetric features on multiple time scales, $S_{temp}$. The $S_{temp}$ would be input into the feature concatenation and classification block along with $S_{spat}$.

\subsubsection{Feature classification block}
Through the previous procedures, we obtain two groups of features, i.e., $S_{spat}$ and $S_{temp}$.	These features are first concatenated to a integrated vector $S_{cat}$, which is called final feature map. The final feature map will be used to visualize the model in the following analysis. Through a dropout with the rate of 0.7, the feature is input into a FC layer with two neurons. The output feature is regarded as the possibility($P(c|D), c= 0,1 $) that the EEG data $D$ belongs to each class. The predicted label is that of the class which has maximal possibility. The procedure could be described as:

\begin{equation}
	{y_{pred} = \mathop{\arg\max}_{c} P(c|D), c= 0,1.} \label{classifier}
\end{equation}
where the \emph{$P(c|D)$} is the possibility of \emph{D} belonging to the \emph{c}-th class.

\subsection{Baseline methods}
To verify the performance of our model, we compared the MSBAM with ten baseline methods, which were evaluated on the DEAP or DREAMER dataset. These methods are introduced briefly as follows:

%

\textbf{The method of Liu-2019.} Liu et al. introduced a method named deep canonical correlation analysis(DCCA)~\cite{liu2019multimodal}. The raw EEG signals and peripheral physiological features were transformed by different nonlinear networks to obtain the two groups of features. These features were fused by a weighted sum method after being regularized with the traditional CCA method. Then, the fusion features were used to train a linear SVM classifier for emotion recognition. Evaluated on DEAP, the DCCA method achieved accuracies of 84.33\% and 85.62\% on arousal and valance dimensions.

\textbf{The method of Qiu-2018.} Qiu et al. proposed a multimodal emotion recognition method named correlated attention network (CAN)~\cite{qiu2018correlated}. They extracted features with two Bidirectional Gated Recurrent Unit neural networks, and applied a canonical correlation analysis to calculate the correlation. In the end, the attention mechanism was utilized to extract the features that are important to represent emotional states. Finally, a softmax layer is applied to complete the emotion classification task. Their method achieved accuracies of 84.79\% and 86.45\% on the arousal and valance dimensions of DEAP dataset.

\textbf{The method of Yin-2021.} Yin et al. proposed an ECL\-GCNN model that fused LSTM and GCNN for emotion classification~\cite{YIN2021106954}. In the GCNN, the EEG channels and functional connections between two channels were denoted as the vertex nodes and edges, respectively. The greater value of the edge means the closer relationship between two channels. The features extracted by the GCNNs were input into LSTM networks to extract the higher level features for emotion classification tasks. They attained the accuracies of 90.45\% and 90.60\% on the valance and arousal dimensions of DEAP dataset.

\textbf{The method of Yang-2018.} Yang et al. introduced a pre-processing method for baseline correction, and validated it with their new model termed as parallel convolutional recurrent neural network (PCRNN)~\cite{2018Emotion}. In this model, the pre-processed 2D EEG signals were transformed into a 3D representation according to the spatial topology of the electrodes. A group of CNNs and LSTMs were then utilized to extract the spatial and temporal features, respectively. To classify the emotion states, these features were concatenated and input into a fully connected layer. This method yields accuracies of 90.80\% and 91.03\% on the valance and arousal dimensions of DEAP dataset.

\textbf{The method of Liao-2020.} Liao et al. proposed a multimodal emotion recognition method~\cite{liao2020multimodal}. They transformed the raw EEG signal into 3D representation as in the studied by Yang et al~\cite{2018Emotion}. 2D-CNN kernels were employed to extract the spatial features of EEG signals, and LSTM were utilized to extract the temporal features of peripheral physiological signals. The concatenated spatial and temporal features were fed into a softmax classifier to predict the emotion states. This method achieved accuracies of 91.95\% and 93.06\% on the valance and arousal dimensions of DEAP dataset.

\textbf{The method of Ma-2019.} Ma et al. proposed a method named MMResLSTM based on EEG and peripheral physiological signals~\cite{ma2019emotion}. MMResLSTM contained four LSTMs, and the last three of them had a residual structure. Multimodal data were respectively fed into two MMResLSTM modules that shared parameters to extract the high-level features, and then concatenated to predict the emotion state by a softmax layer. MMResLSTM attained accuracies of 92.30\% and 92.87\% for valance and arousal of DEAP dataset.

\textbf{The method of Huang-2021.} Huang et al. released a method based on the discrepancy of emotional response between two hemispheres, which was named bi-hemisphere discrepancy convolutional neural network (BiDCNN)~\cite{2021Differences}. Three different matrices were constructed and input into the BiDCNN to extract the spatial and temporal features, including the discrepancy features of emotional responses between left and right hemispheres. In the end, all the features were concatenated and fed into a series of Conv layers and FC layers to predict the emotion state. Evaluated on DEAP, Their method achieved accuracies of 94.38\% and 94.72\% on arousal and valance dimensions.

\textbf{The method of Li-2021.} Li et al. proposed a method named dilated fully convolutional networks (DFCN)~\cite{9321553}. They filtered the raw data into four frequency bands, and calculated three kind of features, i.e., Kurtosis feature, Power feature, and DE feature, in each frequency band. Those features were rearranged to a 3D representation. The DFCN contains two convolution layers, three dilated convolution layers and two linear layers. Besides, they introduced Spectral Norm Regularization (SNR) to reduce the sensitivity of distribution. This method yields 94.59\%, 95.32\%, 94.78\% and 95.19\% accuracies on the valance arousal dominance and liking dimensions of DEAP dataset, and achieved accuracies of 93.15\%, 91.30\% and 92.04\% on the valance, arousal and dominance dimensions of DREAMER dataset.

\textbf{The method of Cui-2020.} Cui et al. introduced a method named RACNN~\cite{cui2020eeg}. They employed continuous 1-D convolution to learn temporal representations. Then two parallel branches were implemented. The first branch was designed to capture the regional information between adjacent channels, and the second one was to capture the discriminative feature between the two hemispheres of the brain. At last, they concatenated the features to recognize the emotion state. They conducted the experiments on valance and arousal dimensions of both DEAP and DREAMER datasets, and yields average accuracies of 96.65\%/97.11\% and 95.55\%/97.01\% respectively.

\subsection{Model implementation}
For the MSBAM, the cross-entropy was employed as the loss function. Adam optimizer was utilized to minimize the loss function with 0.001 learning rate initially. The experiment performed 50 epochs, and the learning rate was reduced to 0.0001 at the 30th epoch. During implementing the MSBAM model, the number of branches was set to 2, and the parameters of  time scales ($L_i$) in the three branches in temporal feature extractor were set to 128 and 64, respectively.

\begin{figure}[ht]
	\centering
	\includegraphics[scale=0.22]{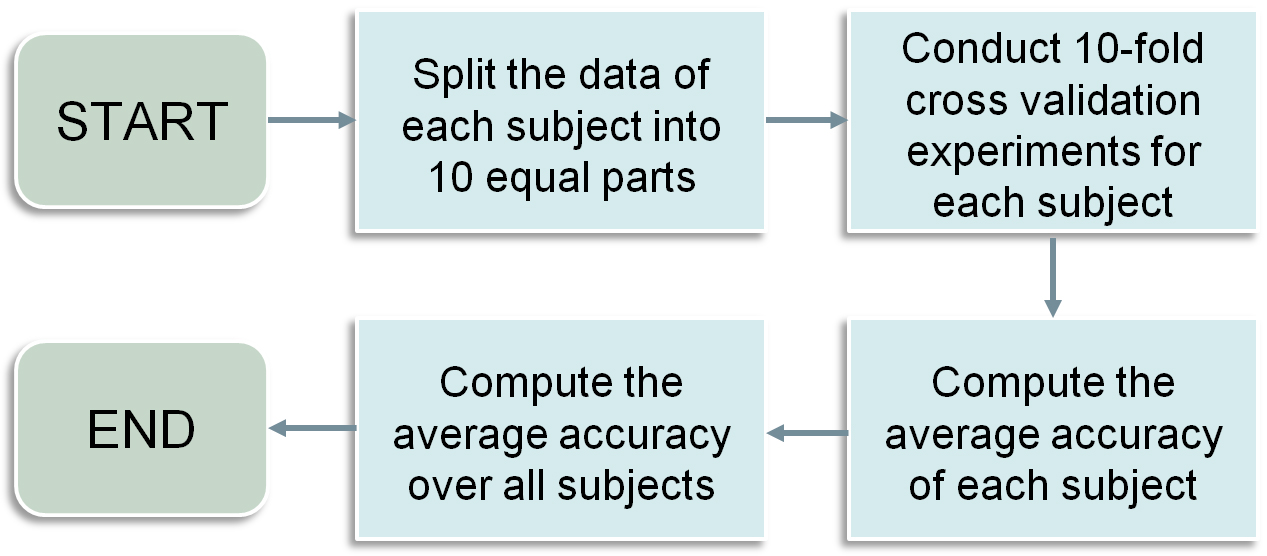}
	\caption{The experimental steps for the classification and evaluation in each dimension.}
	\label{flow-chart}
\end{figure}

The experiments were independently conducted in each emotion dimension of the two datasets. The average accuracy in each dimension was used to evaluate the performance of MSBAM and the compared baseline models. The experimental steps are described as follows, and the flow-chart is illustrated in Fig.~\ref{flow-chart}
\begin{itemize}
	\item [$\bullet$] Randomly split the EEG segments of each subjects into 10 equal parts.
	\item [$\bullet$] Implement 10-fold cross validation experiments for each subject based on the data generated in the above step, and obtain the accuracy of $i$-th subjects, denotes as $Acc_i^1$,  $Acc_i^2$, ..., $Acc_i^{10}$.
	\item [$\bullet$] Compute the average accuracy of each subject as $Acc_i$ = $\frac{1}{10} * \sum_{k = 1}^{10}Acc_i^k$ .
	\item [$\bullet$] Compute the average accuracy over all subjects as $Acc$ = $\frac{1}{N} * \sum_{j = 1}^{N}Acc_j$, where $N$ is the number of subjects in the dataset.
\end{itemize}

\section{Results}

\begin{table*}[!htb]
	\caption{Accuracies of the MSBAM and other baseline methods (mean $\pm$ std.) of DEAP dataset. The symbol\emph{"/"} indicates values of methods are not provided in original studies.}
	\renewcommand{\arraystretch}{1.2}
	\label{t_acc}
	\centering
		
		\begin{tabular}{p{2.8cm} p{2.4cm} p{2.4cm} p{2.4cm} p{2.4cm}}
			\hline
			\hline
			Methods & Valence(\%) & Arousal(\%) & Dominance(\%) & Liking(\%)\\
			\hline
			Tang-2017 & $83.82\pm5.01$ & $83.23\pm2.61$ & $/$ & $/$\\
			Liu-2016 & $85.20\pm/$ & $80.50\pm/$ & $84.90\pm/$ & $82.40\pm/$\\
			Liu-2019 & $85.62\pm3.48$ & $84.33\pm2.25$ & $90.67\pm4.33$ & $/$ \\
			Qiu-2018 & $86.45\pm/$ & $84.79\pm/$ & $/$ & $/$\\
			Yin-2021 & $90.45\pm3.09$ & $90.60\pm2.62$ & $/$ & $/$\\
			Yang-2018 & $90.80\pm3.08$ & $91.03\pm2.99$ & $/$ & $/$\\
			Liao-2020 & $91.95\pm/$ & $93.06\pm/$ & $/$ & $/$\\
			Ma-2019 & $92.30\pm1.55$ & $92.87\pm2.11$ & $/$ & $/$\\
			Huang-2021 & $94.38\pm2.61$ & $94.72\pm2.56$ & $/$ & $/$\\
			Li-2021 & $94.59\pm/$ & $95.32\pm/$  & $94.78\pm/$  & $95.19\pm/$ \\
			Cui-2020 & $96.65\pm2.65$ & $97.11\pm2.01$  & $/$  & $/$ \\
			\textbf{MSBAM} & \bm{$99.36\pm0.46$} & \bm{$99.37\pm0.43$}  & \bm{$99.39\pm0.41$}  & \bm{$99.46\pm0.46$} \\
			\hline
			\hline
		\end{tabular}
\end{table*}
\begin{table}[!htb]
	\caption{Accuracies of the MSBAM and other baseline methods (mean $\pm$ std.) of DREAMER dataset. The symbol\emph{"/"} indicates values of methods are not provided in original studies. The sign '*' indicates that the results reproduced according to the description in the original papers.}
	\renewcommand{\arraystretch}{1.2}
	\label{t_acc2}
	\centering
	\scalebox{0.85}{
		
		\begin{tabular}{p{2cm} p{1.8cm} p{1.8cm} p{2cm}}
			\hline
			\hline
			Methods & Valence(\%) & Arousal(\%) & Dominance(\%)\\
			\hline
			
			Li-2021 & $93.15\pm/$ & $91.30\pm/$  & $92.04\pm/$ \\
			Cui-2020 & $95.55\pm2.18$ & $97.01\pm2.74$  & $/$\\
			Huang-2021$^*$ & $98.35 \pm 0.87$ & $98.66 \pm 1.46$ & $99.01 \pm 0.96$\\
			Liao-2020$^*$ & $98.65 \pm 0.65$ & $98.82 \pm 0.97$ & $99.07 \pm 0.82$\\
			\textbf{MSBAM} & \bm{$99.69\pm0.24$} & \bm{$99.76\pm0.20$}  & \bm{$99.79\pm0.22$}\\
			\hline
			\hline
		\end{tabular}
	}
\end{table}

We respectively conducted the experiments in the dimensions of \emph{valence}, \emph{arousal}, \emph{dominance} and \emph{liking} for the DEAP dataset, and \emph{valence}, \emph{arousal}, \emph{dominance} for DREAMER dataset with MSBAM and the baseline methods. The experimental results of all methods are summarized in Table~\ref {t_acc} and Table~\ref{t_acc2}. In Table 2, the symbol "*" denotes the result reproduced by us, due to the results on current dataset are not provided in the original paper (similarly hereinafter). We used the hyper-parameters provided by the original paper and empirically set parameters that were not provided when we reproduced the model. We could find that our MSBAM yields better performance than the baseline methods. Compared with the best baseline model, MSBAM could improve average recognition accuracy by 2.71\%, 2.26\%, 4.61\% and 4.27\% in \emph{valence}, \emph{arousal}, \emph{dominance} and \emph{liking} dimensions of DEAP, and 1.04\%, 0.94\% and 0.72\% in \emph{valence}, \emph{arousal} and \emph{dominance} dimensions of DREAMER, respectively. In addition, MSBAM achieves smaller standard deviation than the best baseline method. It indicates that MSBAM has better robustness than the baseline methods.
These experiments were conducted to verify the effectiveness of the MSBAM, and compare its performance with the baseline methods.
\begin{figure}[!htb]
	\centering
	\subfigure[]{\includegraphics[scale=1]{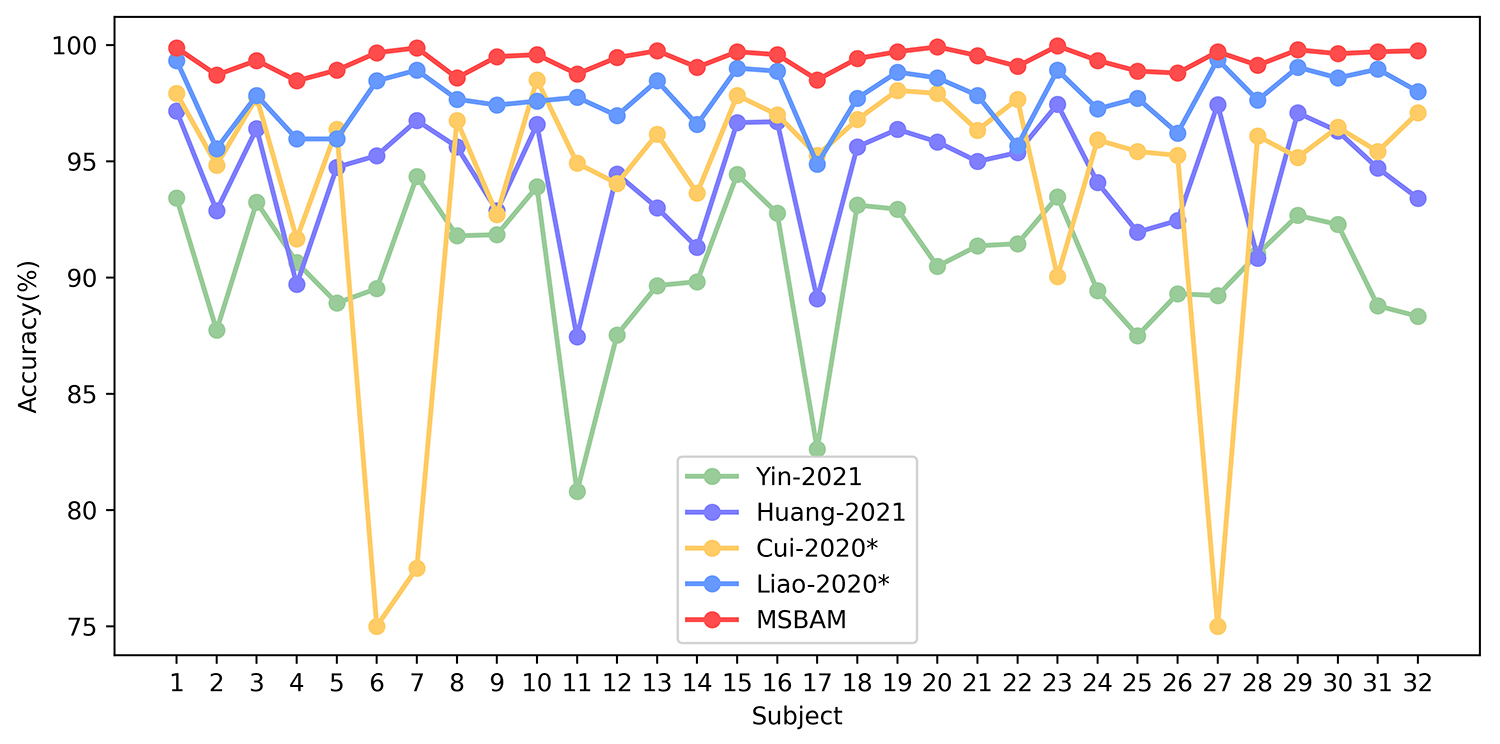}
		\label{S-V}}
	\hfil
	\subfigure[]{\includegraphics[scale=1]{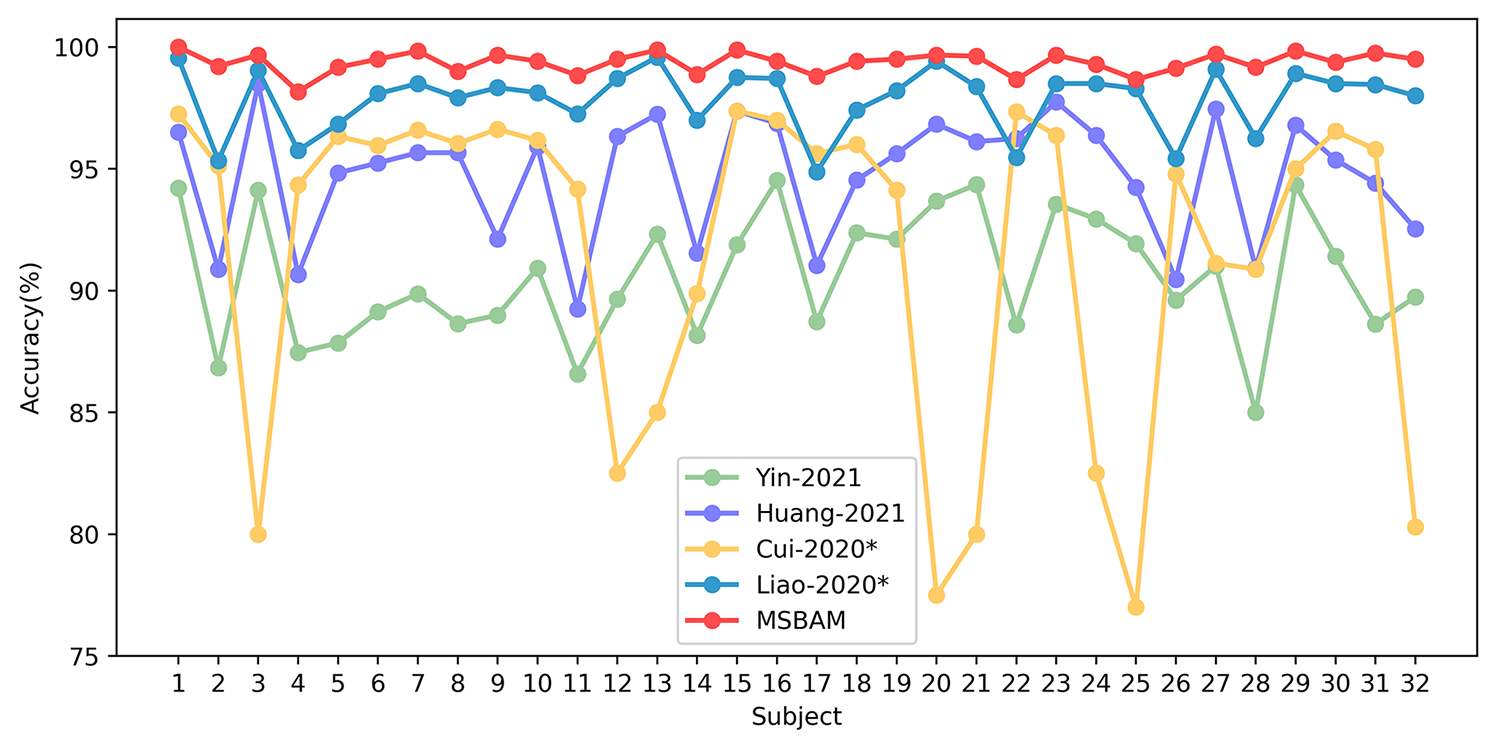}
		\label{S-A}}
	\caption{Average accuracies for every subject on \emph{Valance}(a) and \emph{Arousal}(b) dimension.}
	\label{fig_sim}
\end{figure}

\begin{figure*}[!htb]
	\centering
	\includegraphics[scale=0.23]{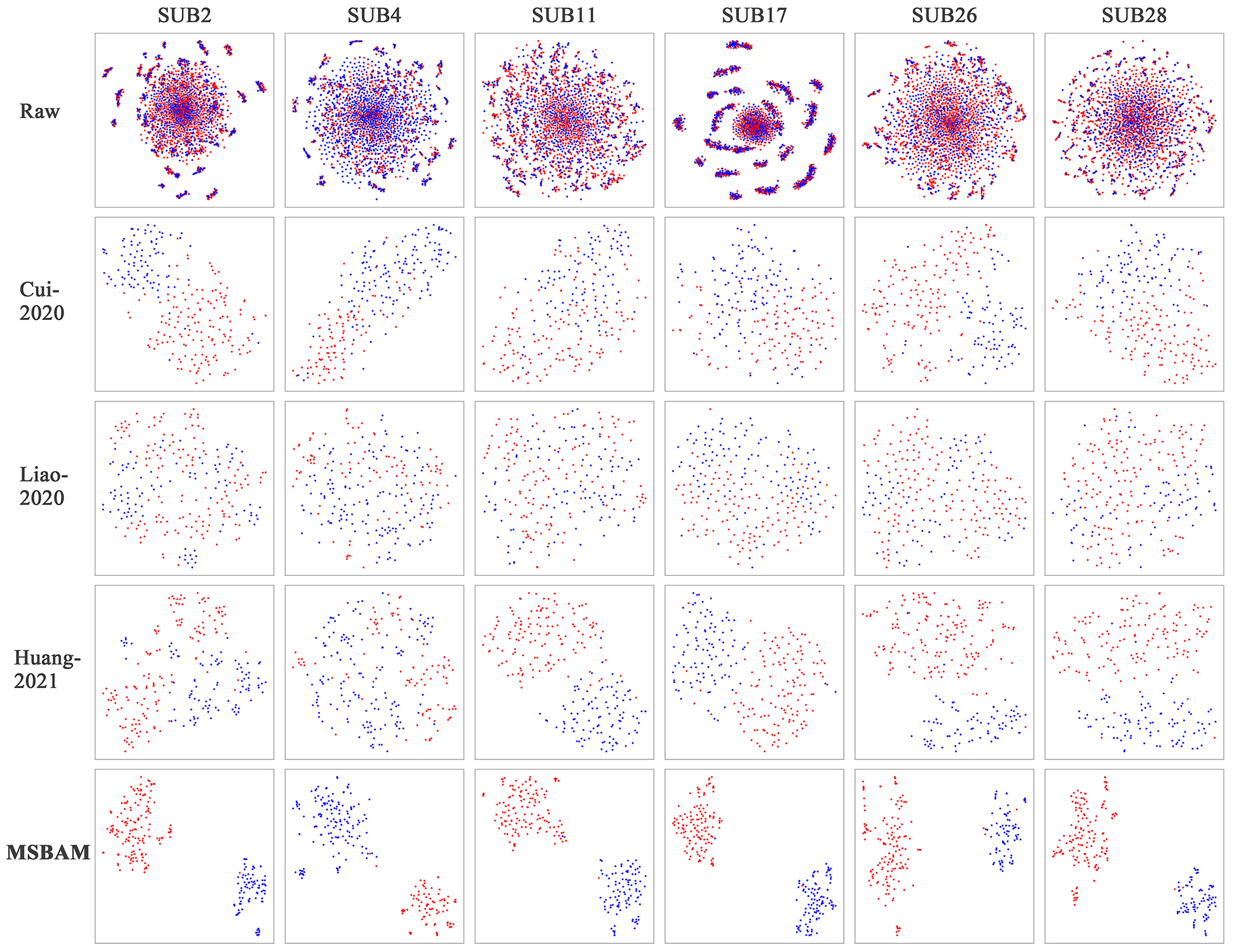}
	\caption{Visualization of the final feature maps using $t$-SNE. The final feature maps are features before the last FC layer(e.g. the $S_{cat}$ in the proposed MSBAM).}
	\label{t-SNE}
\end{figure*}

To observe the performance of proposed MSBAM of each subject, two line charts are presented in Fig. \ref{fig_sim}. In the charts, we present the results on DEAP dataset from four baseline methods, i.e., $Cui-2020$ and $Liao-2020$, $Yin-2021$ and $Huang-2021$. We could find that all the methods yield analogical tendencies as the subject changes. The MSBAM shows better performance than other models. It could yield the better accuracies on all subjects.

To further investigate the results of MSBAM, the features before the last FC layer(final feature map) in \emph{Valence} dimension of DEAP dataset were visualized by the $t$-SNE~\cite{van2008visualizing}. In this work, we reproduced three baseline methods to compare with the proposed MSBAM. Six subjects with lowest accuracies were selected. The features of $t$-SNE are shown in Fig.~\ref{t-SNE}. The first row in the figure illustrated the features obtained from the original data. The second to the last row presented the final feature maps of test set in the first fold of each subject. We could observe that the features extracted in the proposed MSBAM have better intra-category similarity and inter-category separability than the baseline method, which can facilitate the classification.

\section{Discussions}

In current study, we proposed a method MSBAM based on the physiological mechanisms of emotion and the characteristics of EEG to classify the two-class classification in four emotional dimensions, i.e., arousal, valence, dominance and liking, respectively. In each dimension, the EEG data was assigned a low-level or high-level states based on a threshold of the self-rating scores. The 10-fold cross validation was used in intra-subject classification experiments.

The experimental results demonstrate that the proposed MSBAM model outperforms the compared baseline methods. This may be attributed to the model structure that extracts the features of multi-scale and bi-hemispheric asymmetry, simultaneously. To further verify the rationality of the MSBAM, several ablation experiments were conducted on the two datasets. In the following discussion, we will take the DEAP dataset as an example. First, we removed all \emph{FNL} in the model to verify the effectiveness of the normalization method before feature fusion. Then, we further checked whether the spatial feature extractor and two branches in temporal feature extractor block were indispensable. The results of several ablation experiments are shown in Table~\ref{ablation}. MSBAM-r\emph{i} means that the \emph{i}-th branch in the temporal feature extractor block was removed, and the MSBAM-rSpat means the spatial feature extractor was removed, in the original MSBAM. 
	
As shown in Table~\ref{ablation}, we could find that FNL played an important role in the model. One possible explanation is that the FNL is able to scale the features of each branch into the same shape and value domain. Because of the discrepancy of convolution kernels, the feature maps of each branch have different length and value range. These feature maps were transformed to 40 features with the range of 0 to 1 after being processed by the FNL. This procedure may reduce the influence of the weight bias introduced by the discrepancy. In brief, removing all parts of the branches in temporal feature extractor block, or the spatial feature extractor block will reduce the accuracies obtained by the original MSBAM. These results indicate that MSBAM performs better than its variant.

To further evaluate the performance of MSBAM, we further conducted the multiple classification tasks of the four categories of emotions, i.e., high valence and high arousal, high valence and low arousal, low valence and high arousal, and low valence and low arousal, as in previous study~\cite{Cao2020EEG}. The MSBAM yielded the accuracy of $99.20 \pm 0.86$. 

%

\begin{table*}[!htb]
	\caption{Average accuracies(\%) of the multi-scales ablation experiments(mean $\pm$ std.).}
	\renewcommand{\arraystretch}{1.2}
	\label{ablation}
	\centering
	\scalebox{1}{
		
		\begin{tabular}{p{3cm} p{2.2cm} p{2.2cm} p{2.2cm} p{2.2cm} p{2.2cm}}
			\hline
			\hline
			Methods & Valence(\%) & Arousal(\%) & Dominance(\%) & Liking(\%) \\
			\hline		
			MSBAM-rFNL & $98.16 \pm 0.89$&$98.23 \pm 0.95$&$98.32 \pm 0.98$&$98.38 \pm 1.01$ \\
			MSBAM-r1-r2 & $97.33 \pm 1.54$&$95.59 \pm 3.45$&$96.20 \pm 3.02$&$95.68 \pm 2.52$ \\
			MSBAM-r2-rSpat & $98.11 \pm 1.64$&$96.84 \pm 2.62$&$97.12 \pm 2.59$&$96.43 \pm 2.52$ \\			
			MSBAM-r1-rSpat & $98.17 \pm 1.45$&$96.32 \pm 3.71$&$96.85 \pm 2.99$&$96.46 \pm 2.44$ \\			
			MSBAM-r1 & $99.20 \pm 0.53$&$99.08 \pm 0.56$&$98.93 \pm 0.88$&$99.12 \pm 0.53$ \\
			MSBAM-r2 & $99.16 \pm 0.57$&$99.09 \pm 0.55$&$98.97 \pm 0.84$&$99.09 \pm 0.56$ \\				
			MSBAM-rSpat & $99.32 \pm 0.50$&$99.26 \pm 0.48$&$99.17 \pm 0.61$&$99.27 \pm 0.52$ \\				
			\textbf{MSBAM} & \bm{$99.36\pm0.46$} & \bm{$99.37\pm0.43$}  & \bm{$99.39\pm0.41$}  & \bm{$99.46\pm0.46$} \\		
			\hline
			\hline
		\end{tabular}
	}
\end{table*}

Although the MSBAM shows better performance than all the baseline methods, some limitations should be mentioned. In the Table~\ref {t_acc}, we could find that all methods can obtain accuracy above 80\%. However, there exists a problem of data leakage because of the approach to obtaining the training data and testing data for all the trial-dependent method, which is a common phenomenon in existing DL models~\cite{ding2021tsception, 10.1007/978-981-16-2336-3_3}. A few studies try to evaluate the DL models on more challenging scenarios, trial-independent classification within subject~\cite{ding2021tsception, 10.1007/978-981-16-2336-3_3} and subject-independent classification~\cite{HU2021177, zhao2021plug}. Those two scenarios should be adopted in future studies when we evaluate the DL models to avoid data leakage and biased evaluation.

\section{Conclusion}

In this paper, we propose an emotion recognition method MSBAM to extract the multi-scale and bi-hemispheric asymmetric EEG features for emotion classification. The proposed MSBAM could achi\-eve average accuracies of 99.36\%, 99.37\%, 99.39\% and 99.46\% in \emph{Valence}, \emph{Arousal}, \emph{Dominance} and \emph{Liking} dimensions on DEAP dataset and 99.69\%, 99.76\% and 99.79\% in \emph{Valence}, \emph{Arousal}, and \emph{Dominance} dimensions on DREA\-MER dataset, respectively, which outperformed all the baseline methods in the subject-dependent classification scenario. Although some limitations should be addressed in future studies, current study demonstrates that exploring the multi-scale features and utilizing the neural mechanism of the emotions such as the bi-hemispheric asymmetry to design the DL model, are beneficial for emotion recognition.

\section*{Acknowledgments}
We sincerely thank all the reviewers for their insightful comments and constructive suggestions. This work was supported in part by the National Natural Science Foundation of China under Grant No.62076209.


\bibliographystyle{cas-model2-names}


\begin{thebibliography}{41}
	\expandafter\ifx\csname natexlab\endcsname\relax\def\natexlab#1{#1}\fi
	\providecommand{\url}[1]{\texttt{#1}}
	\providecommand{\href}[2]{#2}
	\providecommand{\path}[1]{#1}
	\providecommand{\DOIprefix}{doi:}
	\providecommand{\ArXivprefix}{arXiv:}
	\providecommand{\URLprefix}{URL: }
	\providecommand{\Pubmedprefix}{pmid:}
	\providecommand{\doi}[1]{\href{http://dx.doi.org/#1}{\path{#1}}}
	\providecommand{\Pubmed}[1]{\href{pmid:#1}{\path{#1}}}
	\providecommand{\bibinfo}[2]{#2}
	\ifx\xfnm\relax \def\xfnm[#1]{\unskip,\space#1}\fi
	\bibitem[{Aydın(2020)}]{8933102}
	\bibinfo{author}{Aydın, S.}, \bibinfo{year}{2020}.
	\newblock \bibinfo{title}{Deep learning classification of neuro-emotional phase
		domain complexity levels induced by affective video film clips}.
	\newblock \bibinfo{journal}{IEEE Journal of Biomedical and Health Informatics}
	\bibinfo{volume}{24}, \bibinfo{pages}{1695--1702}.
	\bibitem[{Cabanac(2002)}]{2003What}
	\bibinfo{author}{Cabanac, M.}, \bibinfo{year}{2002}.
	\newblock \bibinfo{title}{What is emotion?}
	\newblock \bibinfo{journal}{Behavioural Processes} \bibinfo{volume}{60},
	\bibinfo{pages}{69--83}.
	\bibitem[{Cao et~al.(2020)Cao, Hao, Wang, Gao and Xiang}]{Cao2020EEG}
	\bibinfo{author}{Cao, R.}, \bibinfo{author}{Hao, Y.}, \bibinfo{author}{Wang,
		X.}, \bibinfo{author}{Gao, Y.}, \bibinfo{author}{Xiang, J.},
	\bibinfo{year}{2020}.
	\newblock \bibinfo{title}{{EEG} functional connectivity underlying emotional
		valance and arousal using minimum spanning trees}.
	\newblock \bibinfo{journal}{Frontiers in Neuroscience} \bibinfo{volume}{14},
	\bibinfo{pages}{355}.
	\bibitem[{Cui et~al.(2020)Cui, Liu, Zhang, Chen, Wang and Chen}]{cui2020eeg}
	\bibinfo{author}{Cui, H.}, \bibinfo{author}{Liu, A.}, \bibinfo{author}{Zhang,
		X.}, \bibinfo{author}{Chen, X.}, \bibinfo{author}{Wang, K.},
	\bibinfo{author}{Chen, X.}, \bibinfo{year}{2020}.
	\newblock \bibinfo{title}{{EEG}-based emotion recognition using an end-to-end
		regional-asymmetric convolutional neural network}.
	\newblock \bibinfo{journal}{Knowledge-Based Systems} \bibinfo{volume}{205},
	\bibinfo{pages}{106243}.
	\bibitem[{Ding et~al.(2022)Ding, Robinson, Zhang, Zeng and
		Guan}]{ding2021tsception}
	\bibinfo{author}{Ding, Y.}, \bibinfo{author}{Robinson, N.},
	\bibinfo{author}{Zhang, S.}, \bibinfo{author}{Zeng, Q.},
	\bibinfo{author}{Guan, C.}, \bibinfo{year}{2022}.
	\newblock \bibinfo{title}{Tsception: Capturing temporal dynamics and spatial
		asymmetry from {EEG} for emotion recognition}.
	\newblock \bibinfo{journal}{IEEE Transactions on Affective Computing} ,
	\DOIprefix\doi{10.1109/TAFFC.2022.3169001}.
	\bibitem[{Dolan(2002)}]{Dolan1191}
	\bibinfo{author}{Dolan, R.J.}, \bibinfo{year}{2002}.
	\newblock \bibinfo{title}{Emotion, cognition, and behavior}.
	\newblock \bibinfo{journal}{Science} \bibinfo{volume}{298},
	\bibinfo{pages}{1191--1194}.
	\bibitem[{{El Ayadi} et~al.(2011){El Ayadi}, Kamel and Karray}]{ELAYADI2011}
	\bibinfo{author}{{El Ayadi}, M.}, \bibinfo{author}{Kamel, M.S.},
	\bibinfo{author}{Karray, F.}, \bibinfo{year}{2011}.
	\newblock \bibinfo{title}{Survey on speech emotion recognition: Features,
		classification schemes, and databases}.
	\newblock \bibinfo{journal}{Pattern Recognition} \bibinfo{volume}{44},
	\bibinfo{pages}{572--587}.
	\bibitem[{Hu et~al.(2021)Hu, Wang, Jia, Bu, Sutcliffe and Feng}]{HU2021177}
	\bibinfo{author}{Hu, J.}, \bibinfo{author}{Wang, C.}, \bibinfo{author}{Jia,
		Q.}, \bibinfo{author}{Bu, Q.}, \bibinfo{author}{Sutcliffe, R.},
	\bibinfo{author}{Feng, J.}, \bibinfo{year}{2021}.
	\newblock \bibinfo{title}{{ScalingNet}: Extracting features from raw {EEG} data
		for emotion recognition}.
	\newblock \bibinfo{journal}{Neurocomputing} \bibinfo{volume}{463},
	\bibinfo{pages}{177--184}.
	\bibitem[{Hua\-ng et~al.(2021)Hua\-ng, Chen, Liu, Zheng, Tian and
		Jiang}]{2021Differences}
	\bibinfo{author}{Hua\-ng, D.}, \bibinfo{author}{Chen, S.},
	\bibinfo{author}{Liu, C.}, \bibinfo{author}{Zheng, L.},
	\bibinfo{author}{Tian, Z.}, \bibinfo{author}{Jiang, D.},
	\bibinfo{year}{2021}.
	\newblock \bibinfo{title}{Differences first in asymmetric brain: A
		bi-hemisphere discrepancy convolutional neural network for {EEG} emotion
		recognition}.
	\newblock \bibinfo{journal}{Neurocomputing} \bibinfo{volume}{448},
	\bibinfo{pages}{140--151}.
	\bibitem[{Katsigiannis and Ramzan(2018)}]{7887697}
	\bibinfo{author}{Katsigiannis, S.}, \bibinfo{author}{Ramzan, N.},
	\bibinfo{year}{2018}.
	\newblock \bibinfo{title}{Dreamer: A database for emotion recognition through
		{EEG} and {ECG} signals from wireless low-cost off-the-shelf devices}.
	\newblock \bibinfo{journal}{IEEE Journal of Biomedical and Health Informatics}
	\bibinfo{volume}{22}, \bibinfo{pages}{98--107}.
	\bibitem[{K{\i}l{\i}{\c{c}} and Ayd{\i}n(2022)}]{kilicc2022classification}
	\bibinfo{author}{K{\i}l{\i}{\c{c}}, B.}, \bibinfo{author}{Ayd{\i}n, S.},
	\bibinfo{year}{2022}.
	\newblock \bibinfo{title}{Classification of contrasting discrete emotional
		states indicated by {EEG} based graph theoretical network measures}.
	\newblock \bibinfo{journal}{Neuroinformatics} , \bibinfo{pages}{1--15}.
	\bibitem[{Ko et~al.(2021)Ko, Jeon, Jeong and Suk}]{Ko2021}
	\bibinfo{author}{Ko, W.}, \bibinfo{author}{Jeon, E.}, \bibinfo{author}{Jeong,
		S.}, \bibinfo{author}{Suk, H.I.}, \bibinfo{year}{2021}.
	\newblock \bibinfo{title}{Multi-scale neural network for {EEG} representation
		learning in {BCI}}.
	\newblock \bibinfo{journal}{IEEE Computational Intelligence Magazine}
	\bibinfo{volume}{16}, \bibinfo{pages}{31--45}.
	\bibitem[{Koelstra et~al.(2012)Koelstra, Muhl, Soleymani, Lee, Yazdani,
		Ebrahimi, Pun, Nijholt and Patras}]{Koelstra2012DEAP}
	\bibinfo{author}{Koelstra, S.}, \bibinfo{author}{Muhl, C.},
	\bibinfo{author}{Soleymani, M.}, \bibinfo{author}{Lee, J.S.},
	\bibinfo{author}{Yazdani, A.}, \bibinfo{author}{Ebrahimi, T.},
	\bibinfo{author}{Pun, T.}, \bibinfo{author}{Nijholt, A.},
	\bibinfo{author}{Patras, I.}, \bibinfo{year}{2012}.
	\newblock \bibinfo{title}{{DEAP}: A database for emotion analysis using
		physiological signals}.
	\newblock \bibinfo{journal}{IEEE Transactions on Affective Computing}
	\bibinfo{volume}{3}, \bibinfo{pages}{18--31}.
	\bibitem[{Li et~al.(2021a)Li, Chai, Wang, Yang and Du}]{9321553}
	\bibinfo{author}{Li, D.}, \bibinfo{author}{Chai, B.}, \bibinfo{author}{Wang,
		Z.}, \bibinfo{author}{Yang, H.}, \bibinfo{author}{Du, W.},
	\bibinfo{year}{2021}a.
	\newblock \bibinfo{title}{{EEG} emotion recognition based on {3-D} feature
		representation and dilated fully convolutional networks}.
	\newblock \bibinfo{journal}{IEEE Transactions on Cognitive and Developmental
		Systems} \bibinfo{volume}{13}, \bibinfo{pages}{885--897}.
	\bibitem[{Li et~al.(2020)Li, Xu, Wang, Fang and Ji}]{lidl2020}
	\bibinfo{author}{Li, D.}, \bibinfo{author}{Xu, J.}, \bibinfo{author}{Wang, J.},
	\bibinfo{author}{Fang, X.}, \bibinfo{author}{Ji, Y.}, \bibinfo{year}{2020}.
	\newblock \bibinfo{title}{A multi-scale fusion convolutional neural network
		based on attention mechanism for the visualization analysis of {EEG} signals
		decoding}.
	\newblock \bibinfo{journal}{IEEE Transactions on Neural Systems and
		Rehabilitation Engineering} \bibinfo{volume}{28},
	\bibinfo{pages}{2615--2626}.
	\bibitem[{Li et~al.(2019a)Li, Liu, Si, Li, Li, Zhu, Huang, Zeng, Yao, Zhang and
		Xu}]{Lipy2019}
	\bibinfo{author}{Li, P.}, \bibinfo{author}{Liu, H.}, \bibinfo{author}{Si, Y.},
	\bibinfo{author}{Li, C.}, \bibinfo{author}{Li, F.}, \bibinfo{author}{Zhu,
		X.}, \bibinfo{author}{Huang, X.}, \bibinfo{author}{Zeng, Y.},
	\bibinfo{author}{Yao, D.}, \bibinfo{author}{Zhang, Y.}, \bibinfo{author}{Xu,
		P.}, \bibinfo{year}{2019}a.
	\newblock \bibinfo{title}{{EEG} based emotion recognition by combining
		functional connectivity network and local activations}.
	\newblock \bibinfo{journal}{IEEE Transactions on Biomedical Engineering}
	\bibinfo{volume}{66}, \bibinfo{pages}{2869--2881}.
	\bibitem[{Li et~al.(2021b)Li, Huan, Hou, Tian, Zhang and Song}]{liwei2021}
	\bibinfo{author}{Li, W.}, \bibinfo{author}{Huan, W.}, \bibinfo{author}{Hou,
		B.}, \bibinfo{author}{Tian, Y.}, \bibinfo{author}{Zhang, Z.},
	\bibinfo{author}{Song, A.}, \bibinfo{year}{2021}b.
	\newblock \bibinfo{title}{Can emotion be transferred?-a review on transfer
		learning for {EEG}-based emotion recognition}.
	\newblock \bibinfo{journal}{IEEE Transactions on Cognitive and Developmental
		Systems} , 
	\DOIprefix\doi{10.1109/TCDS.2021.3098842}.
	\bibitem[{Li et~al.(2019b)Li, Zheng, Wang, Zong and Cui}]{2019From}
	\bibinfo{author}{Li, Y.}, \bibinfo{author}{Zheng, W.}, \bibinfo{author}{Wang,
		L.}, \bibinfo{author}{Zong, Y.}, \bibinfo{author}{Cui, Z.},
	\bibinfo{year}{2019}b.
	\newblock \bibinfo{title}{From regional to global brain: A novel hierarchical
		spatial-temporal neural network model for {EEG} emotion recognition}.
	\newblock \bibinfo{journal}{IEEE Transactions on Affective Computing} ,
	\DOIprefix\doi{10.1109/TAFFC.2019.2922912}.
	\bibitem[{Li et~al.(2021c)Li, Zheng, Zong, Cui, Zhang and Zhou}]{2018A}
	\bibinfo{author}{Li, Y.}, \bibinfo{author}{Zheng, W.}, \bibinfo{author}{Zong,
		Y.}, \bibinfo{author}{Cui, Z.}, \bibinfo{author}{Zhang, T.},
	\bibinfo{author}{Zhou, X.}, \bibinfo{year}{2021}c.
	\newblock \bibinfo{title}{A bi-hemisphere domain adversarial neural network
		model for {EEG} emotion recognition}.
	\newblock \bibinfo{journal}{IEEE Transactions on Affective Computing}
	\bibinfo{volume}{12}, \bibinfo{pages}{494--504}.
	\bibitem[{Liao et~al.(2020)Liao, Zhong, Zhu and Cai}]{liao2020multimodal}
	\bibinfo{author}{Liao, J.}, \bibinfo{author}{Zhong, Q.}, \bibinfo{author}{Zhu,
		Y.}, \bibinfo{author}{Cai, D.}, \bibinfo{year}{2020}.
	\newblock \bibinfo{title}{Multimodal physiological signal emotion recognition
		based on convolutional recurrent neural network}.
	\newblock \bibinfo{journal}{{IOP} Conference Series: Materials Science and
		Engineering} \bibinfo{volume}{782}, \bibinfo{pages}{032005}.
	\bibitem[{Liu et~al.(2019)Liu, Qiu, Zheng and Lu}]{liu2019multimodal}
	\bibinfo{author}{Liu, W.}, \bibinfo{author}{Qiu, J.L.}, \bibinfo{author}{Zheng,
		W.L.}, \bibinfo{author}{Lu, B.L.}, \bibinfo{year}{2019}.
	\newblock \bibinfo{title}{Multimodal emotion recognition using deep canonical
		correlation analysis}.
	\newblock \href{http://arxiv.org/abs/1908.05349}{\tt arXiv:1908.05349}.
	\bibitem[{Liu et~al.(2016)Liu, Zheng and Lu}]{liu2016emotion}
	\bibinfo{author}{Liu, W.}, \bibinfo{author}{Zheng, W.L.}, \bibinfo{author}{Lu,
		B.L.}, \bibinfo{year}{2016}.
	\newblock \bibinfo{title}{Emotion recognition using multimodal deep learning},
	in: \bibinfo{editor}{Hirose, A.}, \bibinfo{editor}{Ozawa, S.},
	\bibinfo{editor}{Doya, K.}, \bibinfo{editor}{Ikeda, K.},
	\bibinfo{editor}{Lee, M.}, \bibinfo{editor}{Liu, D.} (Eds.),
	\bibinfo{booktitle}{Neural Information Processing},
	\bibinfo{publisher}{Springer International Publishing},
	\bibinfo{address}{Cham}. pp. \bibinfo{pages}{521--529}.
	\bibitem[{Ma et~al.(2019b)Ma, Tang, Zheng and Lu}]{ma2019emotion}
	\bibinfo{author}{Ma, J.}, \bibinfo{author}{Tang, H.}, \bibinfo{author}{Zheng,
		W.L.}, \bibinfo{author}{Lu, B.L.}, \bibinfo{year}{2019}b.
	\newblock \bibinfo{title}{Emotion recognition using multimodal residual {LSTM}
		network}, in: \bibinfo{booktitle}{Proceedings of the 27th ACM International
		Conference on Multimedia}, \bibinfo{publisher}{Association for Computing
		Machinery}, \bibinfo{address}{New York, NY, USA}. p.
	\bibinfo{pages}{176–183}.
	\bibitem[{Van~der Maaten and Hinton(2008)}]{van2008visualizing}
	\bibinfo{author}{Van~der Maaten, L.}, \bibinfo{author}{Hinton, G.},
	\bibinfo{year}{2008}.
	\newblock \bibinfo{title}{Visualizing data using {t-SNE}.}
	\newblock \bibinfo{journal}{Journal of machine learning research}
	\bibinfo{volume}{9}, \bibinfo{pages}{2579--2605}.
	\bibitem[{Moon et~al.(2020)Moon, Chen, Hsieh, Wang and Lee}]{MOON202096}
	\bibinfo{author}{Moon, S.E.}, \bibinfo{author}{Chen, C.J.},
	\bibinfo{author}{Hsieh, C.J.}, \bibinfo{author}{Wang, J.L.},
	\bibinfo{author}{Lee, J.S.}, \bibinfo{year}{2020}.
	\newblock \bibinfo{title}{Emotional {EEG} classification using connectivity
		features and convolutional neural networks}.
	\newblock \bibinfo{journal}{Neural Networks} \bibinfo{volume}{132},
	\bibinfo{pages}{96--107}.
	\bibitem[{Phan et~al.(2021)Phan, Kim, Yang and Lee}]{phan2021eeg}
	\bibinfo{author}{Phan, T.D.T.}, \bibinfo{author}{Kim, S.H.},
	\bibinfo{author}{Yang, H.J.}, \bibinfo{author}{Lee, G.S.},
	\bibinfo{year}{2021}.
	\newblock \bibinfo{title}{{EEG}-based emotion recognition by convolutional
		neural network with multi-scale kernels}.
	\newblock \bibinfo{journal}{Sensors} \bibinfo{volume}{21}.
	\bibitem[{Qiu et~al.(2018)Qiu, Li and Hu}]{qiu2018correlated}
	\bibinfo{author}{Qiu, J.L.}, \bibinfo{author}{Li, X.Y.}, \bibinfo{author}{Hu,
		K.}, \bibinfo{year}{2018}.
	\newblock \bibinfo{title}{Correlated attention networks for multimodal emotion
		recognition}, in: \bibinfo{booktitle}{2018 IEEE International Conference on
		Bioinformatics and Biomedicine (BIBM)}, pp. \bibinfo{pages}{2656--2660}.
	\bibitem[{Torres et~al.(2020)Torres, Torres, Hern{\'a}ndez-{\'A}lvarez and
		Yoo}]{torres2020eeg}
	\bibinfo{author}{Torres, E.P.}, \bibinfo{author}{Torres, E.A.},
	\bibinfo{author}{Hern{\'a}ndez-{\'A}lvarez, M.}, \bibinfo{author}{Yoo, S.G.},
	\bibinfo{year}{2020}.
	\newblock \bibinfo{title}{{EEG}-based {BCI} emotion recognition: a survey}.
	\newblock \bibinfo{journal}{Sensors} \bibinfo{volume}{20},
	\bibinfo{pages}{5083}.
	\bibitem[{Wang et~al.(2021)Wang, Liu, Qi, Deng and
		Li}]{10.1007/978-981-16-2336-3_3}
	\bibinfo{author}{Wang, H.}, \bibinfo{author}{Liu, K.}, \bibinfo{author}{Qi,
		F.}, \bibinfo{author}{Deng, X.}, \bibinfo{author}{Li, P.},
	\bibinfo{year}{2021}.
	\newblock \bibinfo{title}{{EEG}-based emotion recognition using convolutional
		neural network with functional connections}, in: \bibinfo{editor}{Sun, F.},
	\bibinfo{editor}{Liu, H.}, \bibinfo{editor}{Fang, B.} (Eds.),
	\bibinfo{booktitle}{Cognitive Systems and Signal Processing},
	\bibinfo{publisher}{Springer Singapore}, \bibinfo{address}{Singapore}. pp.
	\bibinfo{pages}{33--40}.
	\bibitem[{Wen et~al.(2017)Wen, Xu and Du}]{2017A}
	\bibinfo{author}{Wen, Z.}, \bibinfo{author}{Xu, R.}, \bibinfo{author}{Du, J.},
	\bibinfo{year}{2017}.
	\newblock \bibinfo{title}{A novel convolutional neural networks for emotion
		recognition based on {EEG} signal}, in: \bibinfo{booktitle}{2017
		International Conference on Security, Pattern Analysis, and Cybernetics
		(SPAC)}, pp. \bibinfo{pages}{672--677}.
	\bibitem[{Yang et~al.(2018a)Yang, Wu, Qiu, Wang and Chen}]{2018Emotion}
	\bibinfo{author}{Yang, Y.}, \bibinfo{author}{Wu, Q.}, \bibinfo{author}{Qiu,
		M.}, \bibinfo{author}{Wang, Y.}, \bibinfo{author}{Chen, X.},
	\bibinfo{year}{2018}a.
	\newblock \bibinfo{title}{Emotion recognition from multi-channel {EEG} through
		parallel convolutional recurrent neural network}, in:
	\bibinfo{booktitle}{2018 International Joint Conference on Neural Networks
		(IJCNN)}, pp. \bibinfo{pages}{1--7}.
	\bibitem[{Yin et~al.(2021a)Yin, Zheng, Hu, Zhang and Cui}]{YIN2021106954}
	\bibinfo{author}{Yin, Y.}, \bibinfo{author}{Zheng, X.}, \bibinfo{author}{Hu,
		B.}, \bibinfo{author}{Zhang, Y.}, \bibinfo{author}{Cui, X.},
	\bibinfo{year}{2021}a.
	\newblock \bibinfo{title}{{EEG} emotion recognition using fusion model of graph
		convolutional neural networks and {LSTM}}.
	\newblock \bibinfo{journal}{Applied Soft Computing} \bibinfo{volume}{100},
	\bibinfo{pages}{106954}.
	\bibitem[{Zhang et~al.(2021)Zhang, Cai, Nie, Xu, Zhao and
		Guan}]{zhang2021atten}
	\bibinfo{author}{Zhang, Y.}, \bibinfo{author}{Cai, H.}, \bibinfo{author}{Nie,
		L.}, \bibinfo{author}{Xu, P.}, \bibinfo{author}{Zhao, S.},
	\bibinfo{author}{Guan, C.}, \bibinfo{year}{2021}.
	\newblock \bibinfo{title}{An end-to-end {3D} convolutional neural network for
		decoding attentive mental state}.
	\newblock \bibinfo{journal}{Neural Networks} \bibinfo{volume}{144},
	\bibinfo{pages}{129--137}.
	\bibitem[{Zhao et~al.(2021a)Zhao, Yan and Lu}]{zhao2021plug}
	\bibinfo{author}{Zhao, L.M.}, \bibinfo{author}{Yan, X.}, \bibinfo{author}{Lu,
		B.L.}, \bibinfo{year}{2021}a.
	\newblock \bibinfo{title}{Plug-and-play domain adaptation for cross-subject
		{EEG}-based emotion recognition}, in: \bibinfo{booktitle}{Proceedings of the
		35th AAAI Conference on Artificial Intelligence}, \bibinfo{organization}{sn}.
	\bibitem[{Zhao et~al.(2021b)Zhao, Tao, Zhang, Xu, Zhang, Hao and
		Chen}]{Zhaosr2021}
	\bibinfo{author}{Zhao, S.}, \bibinfo{author}{Tao, H.}, \bibinfo{author}{Zhang,
		Y.}, \bibinfo{author}{Xu, T.}, \bibinfo{author}{Zhang, K.},
	\bibinfo{author}{Hao, Z.}, \bibinfo{author}{Chen, E.}, \bibinfo{year}{2021}b.
	\newblock \bibinfo{title}{A two-stage 3d {CNN} based learning method for
		spontaneous micro-expression recognition}.
	\newblock \bibinfo{journal}{Neurocomputing} \bibinfo{volume}{448},
	\bibinfo{pages}{276--289}.
	\bibitem[{Zhao et~al.(2019a)Zhao, Zhang, Zhu, You, Kuang and
		Sun}]{zhao2019multi}
	\bibinfo{author}{Zhao, X.}, \bibinfo{author}{Zhang, H.}, \bibinfo{author}{Zhu,
		G.}, \bibinfo{author}{You, F.}, \bibinfo{author}{Kuang, S.},
	\bibinfo{author}{Sun, L.}, \bibinfo{year}{2019}a.
	\newblock \bibinfo{title}{A multi-branch {3D} convolutional neural network for
		{EEG}-based motor imagery classification}.
	\newblock \bibinfo{journal}{IEEE Transactions on Neural Systems and
		Rehabilitation Engineering} \bibinfo{volume}{27},
	\bibinfo{pages}{2164--2177}.
	\bibitem[{Zheng et~al.(2019)Zheng, Zhu and Lu}]{2017Identifying}
	\bibinfo{author}{Zheng, W.L.}, \bibinfo{author}{Zhu, J.Y.},
	\bibinfo{author}{Lu, B.L.}, \bibinfo{year}{2019}.
	\newblock \bibinfo{title}{Identifying stable patterns over time for emotion
		recognition from {EEG}}.
	\newblock \bibinfo{journal}{IEEE Transactions on Affective Computing}
	\bibinfo{volume}{10}, \bibinfo{pages}{417--429}.
\end{thebibliography}

\end{document}